\documentclass[final,5p,times,twocolumn]{article}
\usepackage{graphicx}
\usepackage{authblk}
\usepackage{amssymb}
\usepackage{lineno}
\usepackage{color}
\usepackage{gensymb}

\begin{document}

\newcommand{\red}{\textcolor{black}}
\newcommand{\blue}{\textcolor{black}}
\newcommand{\etc}{\textit{etc.}}
\newcommand{\ie}{\textit{i.e.\,}}
\newcommand{\eg}{\textit{e.g.\,}}
\newcommand{\etal}{\textit{et al.\,\,}}
\newcommand{\roha}{Rohacell\textsuperscript{\textregistered} }
\newcommand{\rohacomma}{Rohacell\textsuperscript{\textregistered}, }
\newcommand{\Tedlar}{Tedlar\textsuperscript{\textregistered} }
\newcommand{\Tedlarcomma}{Tedlar\textsuperscript{\textregistered}, }


\title{\textbf{Characterising encapsulated nuclear waste using cosmic-ray muon tomography}}
\date{}  
\author[1]{A.\,Clarkson}
\author[1]{D.\,J.\,Hamilton}
\author[1]{M.\,Hoek}
\author[1]{D.\,G.\,Ireland}
\author[2]{J.\,R.\,Johnstone}
\author[1]{R.\,Kaiser}
\author[1]{T.\,Keri}
\author[1]{S.\,Lumsden}
\author[1*]{D.\,F.\,Mahon}
\author[1]{B.\,McKinnon}
\author[1]{M.\,Murray}
\author[1]{S.\,Nutbeam-Tuffs}
\author[2]{C.\,Shearer}
\author[1]{G.\,Yang}
\author[2]{C.\,Zimmerman}

\affil[1]{\small{SUPA, School of Physics \& Astronomy, University of Glasgow, Kelvin Building, University Avenue, Glasgow, G12 8QQ, Scotland, UK}}
\affil[2]{\small{National Nuclear Laboratory, Central Laboratory, Sellafield, Seascale, Cumbria, CA20 1PG, England, UK}}
\affil[*]{e-mail: david.mahon@glasgow.ac.uk}

\maketitle

\textbf{Tomographic imaging techniques using the Coulomb scattering of cosmic-ray muons have been shown previously to successfully identify and characterise low- and high-Z materials within an air matrix using a prototype scintillating-fibre tracker system.  Those studies were performed as the first in a series to assess the feasibility of this technology and image reconstruction techniques in characterising the potential high-Z contents of legacy nuclear waste containers for the UK Nuclear Industry.  The present work continues the feasibility study and presents the first images reconstructed from experimental data collected using this small-scale prototype system of low- and high-Z materials encapsulated within a concrete-filled stainless-steel container.   Clear discrimination is observed between the thick steel casing, the concrete matrix and the sample materials assayed.   These reconstructed objects are presented and discussed in detail alongside the implications for future industrial scenarios.}







Muon Tomography (MT) is a burgeoning field of applied scientific investigation. The technique makes use of the penetrating properties of cosmic-ray muons to image the internal composition of large and/or sealed containers that cannot be interrogated with conventional means \eg X-rays.  Since the pioneering experiments in the mid-20th Century led by George~\cite{George1955} and Alvarez~\cite{Alvarez1970} there has been a wide range of applications exploiting muons for imaging purposes.  Recent interest in the field has been sparked within volcanology~\cite{Tanaka2005,Ambrosi2011} and national security for its ability to detect shielded nuclear contraband within large volumes without the need for a manual search~\cite{Borozdin2003a,Gnanvo08,Riggi2013}.     

As cosmic rays impact upon the atmosphere, particles are produced and subsequently decay as they shower towards sea level.  Here, charged muons are detected with a flux in the region of one muon per square centimetre per minute.  These highly-penetrative particles interact with matter via ionising interactions with atomic electrons and via Coulomb scattering from nuclei.  It is this latter mechanism, and its dependence on atomic number Z, that is exploited for MT in this work.

The MT application discussed here is focussed on the identification and characterisation of any remnant nuclear materials stored within legacy nuclear waste containers.  Our recent results published in Refs.~\cite{Clarkson2014a,Clarkson2014b} have shown the potential of locating and characterising high density materials within air using cosmic-ray muon tomographic techniques.  A robust and efficient prototype detector system has been developed along with custom image reconstruction software using probabilistic determination methods that improve on the state-of-the-art.    Simulated results from similar (though of a larger scale) nuclear waste applications have been published previously by Jonkmans \etal~\cite{Jonkmans2013}.

In this paper, first imaging results from this prototype detector are presented for high-Z materials encapsulated within a concrete-filled stainless-steel container.  An overview of the detector system and its operational performance is given in Section~\ref{sec:Detector}.  The test configuration under interrogation is described in Section~\ref{sec:Imaging} with a brief overview of the image reconstruction process.  Results showing the first images reconstructed from experimental data collected with this test configuration are presented in Section~\ref{results} alongside predictions from dedicated simulations.   These are summarised in Section~\ref{summary} with the industrial implications of the results discussed. 

\begin{figure}[t] 
\centering 
\includegraphics[width=0.90\columnwidth,keepaspectratio]{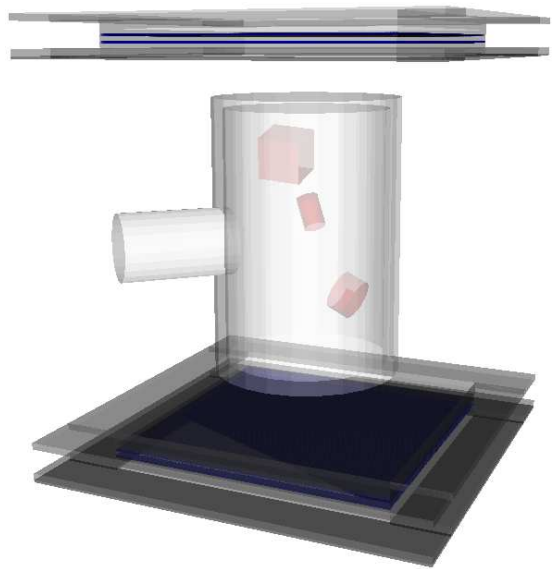}
\caption{\small{\textbf{GEANT4 simulation of the test configuration under investigation.}  An indication of the positions of the lead (top), uranium (centre) and brass (bottom) objects within the actual experimental configuration.  These three objects occupy different planes within the container.  Shown also are the 12\,mm-thick stainless-steel container filled with concrete and the hollow steel mechanical arm that supported the container.  Only the two innermost detector modules are shown.  The orthogonal layers of scintillating fibres and support layers are visible.}}
\label{fig:Geant4}
\end{figure}

\section{Prototype Detector Configuration}\label{sec:Detector}
The design, construction and performance of the small-scale prototype system are all detailed in Ref.~\cite{Clarkson2014a}.  A modular system was chosen based on scintillating fibre and Multi-Anode PhotoMultiplier Tube (MAPMT) technologies.   Four tracking modules, two above and two below the volume under interrogation \ie the assay volume, were constructed.  These consisted of orthogonal layers of 128 Saint Gobain BCF-10 fibres of 2\,mm pitch, chosen based on results from simulation studies outlined in Ref.~\cite{Clarkson2014b}.  Fibres were optically coupled to Hamamatsu H8500 MAPMTs and signals read out by custom data acquisition software.   

\section{Test Configuration \& Image Reconstruction}\label{sec:Imaging}
Previous simulation results presented in Ref.~\cite{Clarkson2014b} showed the anticipated performance of the small-scale prototype system and image reconstruction software in imaging high-Z materials within a cylindrical concrete-filled steel container.  Here, 40\,mm cubic samples of uranium and uranium oxide were simulated using a realistic GEANT4~\cite{geant4} simulation of the detector system that had previously been verified by experimental data.   This study showed clear discrimination between the two high-Z materials and the surrounding concrete matrix.   

To verify these initial predictions with experimental data, a test configuration, shown simulated in Figure~\ref{fig:Geant4}, representing a small-scale industrial waste container was constructed and placed in the centre of the assay volume for interrogation.  This stainless-steel barrel measured 255\,mm in height and 175\,mm in diameter with a wall thickness of 12\,mm (including base and lid).  This thickness was exaggerated compared with the realistic industrial scenario to provide a robust test of the reconstruction ability.  The container was externally supported by a hollow mechanical support arm, 70\,mm in diameter and made of 12\,mm-thick steel, to ensure stability throughout the data collection period.  The steel barrel was filled with concrete and three material samples: a (40\,mm)$^{3}$ cube of lead (with similar density to uranium oxide), a uranium cylinder measuring 30\,mm in length and 20\,mm in diameter, and a brass cylinder of 40\,mm diameter and 20\,mm thick.  These were placed in different planes within the assay volume with the high-Z (low-Z) lead (brass) piece located at the top (bottom) of the container.  

The images shown in Section~\ref{results} were reconstructed from experimental data using custom-developed software based on the Maximum Likelihood Expectation Maximisation algorithm introduced by Schultz \etal in Ref.~\cite{Schultz07} and outlined for this application in Refs.~\cite{Clarkson2014a,Clarkson2014b}.  Data were collected using the constructed prototype system over the course of 30 weeks.  This timescale is not indicative of the required data to provide an accurate determination of the container contents, with first indications of high-Z materials deducible in a much shorter timescale.  Prior to event-by-event data analysis using this software, the assay volume was divided into small volume elements called voxels.  Previous studies presented in Ref.~\cite{Clarkson2014b} showed the size of voxel to influence the minimum achievable image resolution, though smaller voxels require a longer data collection time to produce a reliable image.   Cuboidal voxels of 5\,mm\,x\,5\,mm\,x\,10\,mm were used for all images reconstructed in this work.

The most-likely scattering density, denoted $\lambda$, was reconstructed per voxel in an iterative process.  This parameter is known to exhibit a dependence on the atomic number Z of the scattering material~\cite{Schultz04} and is used as the imaging metric in the results shown in the following section.

\section{Experimental Results}\label{results}
\begin{figure*}[t] 
\centering 
\includegraphics[width=1.0\columnwidth,keepaspectratio]{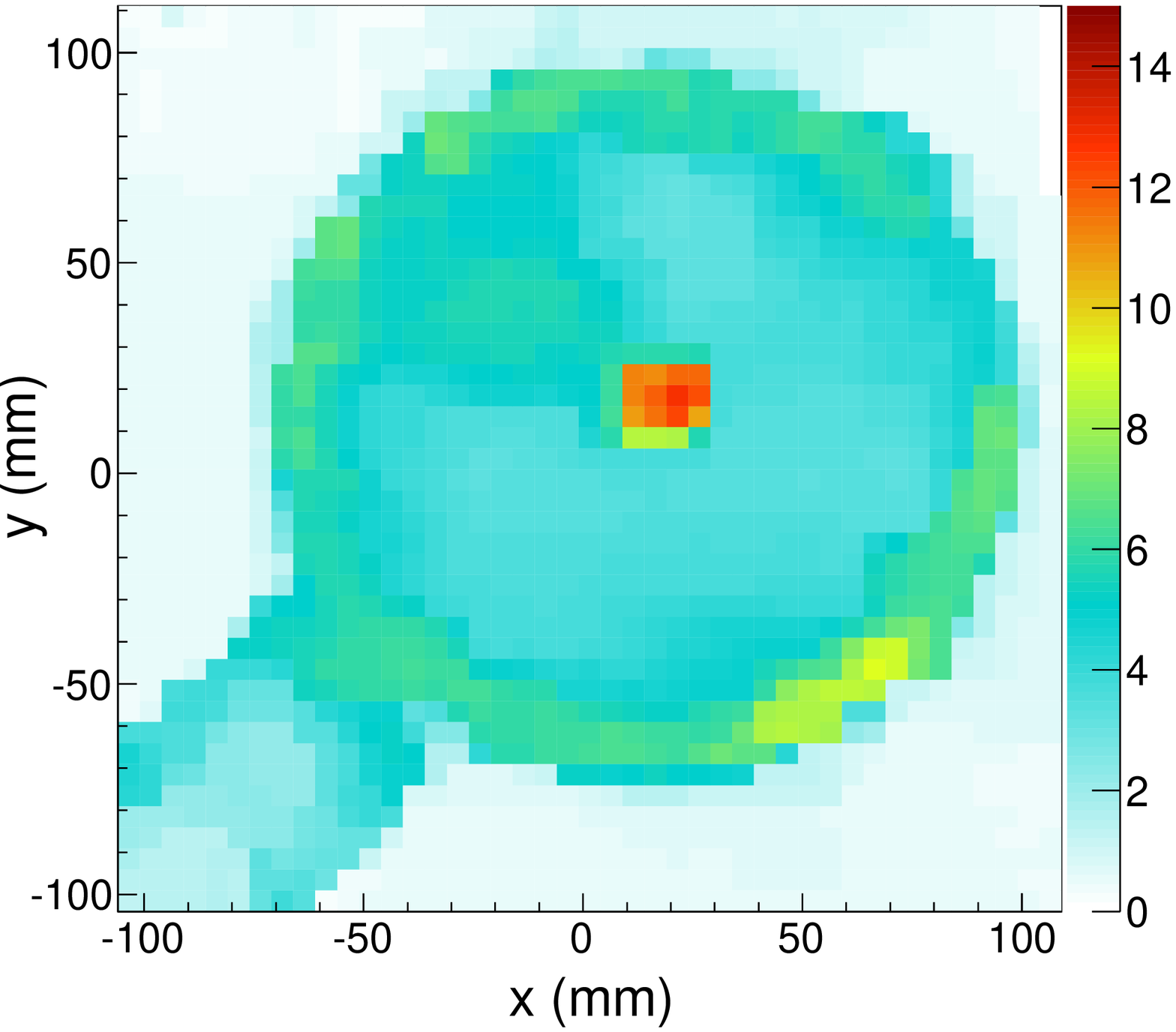}
\includegraphics[width=1.0\columnwidth,keepaspectratio]{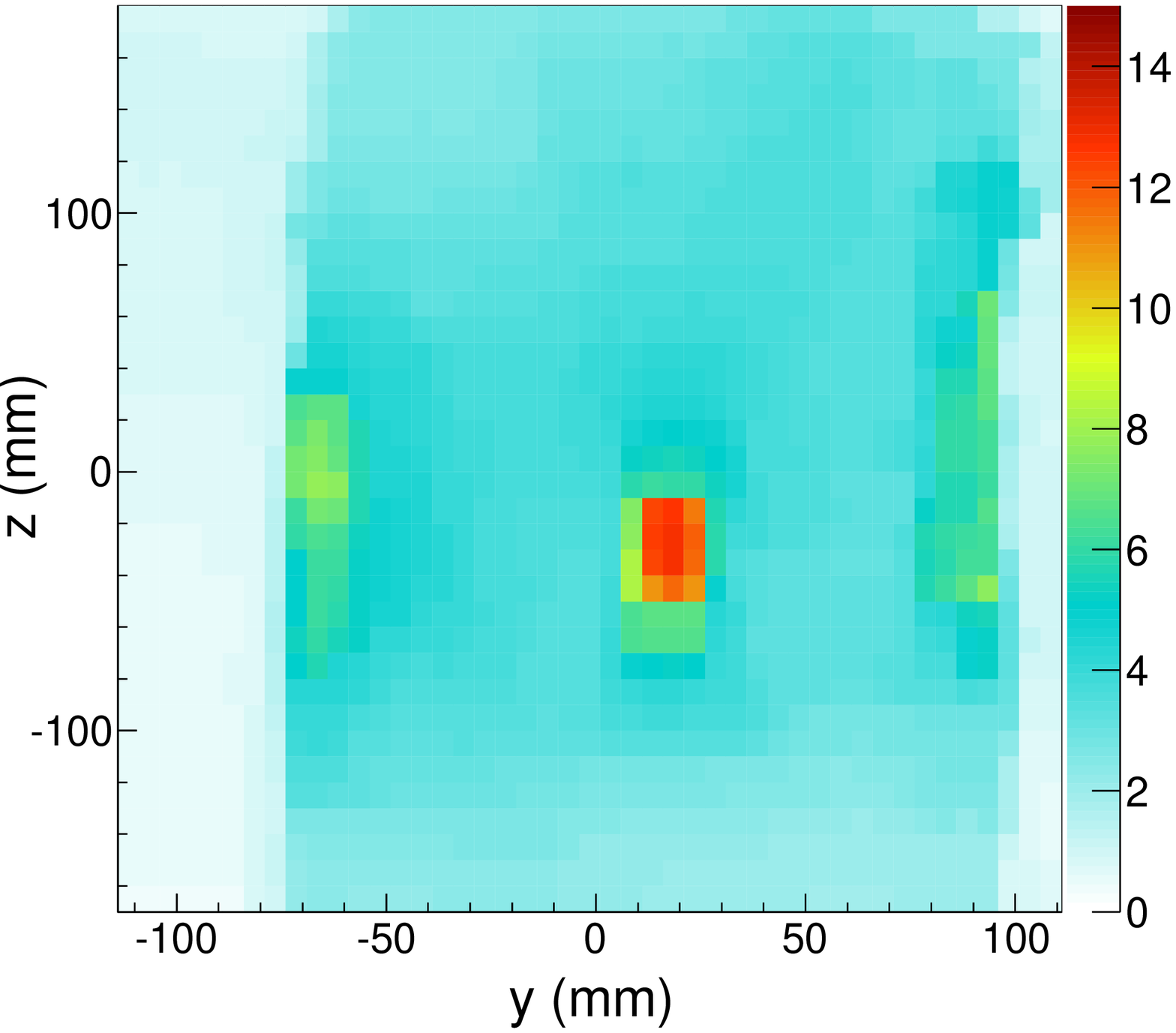}
\caption{\small{\textbf{Tomograms through the uranium-containing region of the barrel.}  Images reconstructed from several weeks of exposure to cosmic-ray muons using the small-scale prototype system.  Shown are a 10\,mm slice in the xy-plane \ie horizontal through the assay volume (left) and a 5\,mm slice in the vertical $yz$-plane (right) through the region containing the uranium cylinder.  Here, the colour scale denotes the most-likely $\lambda$ value in each voxel reconstructed by the image reconstruction software.  Only voxels within the active volume are shown.  The small cylinder of uranium, stainless steel casing, hollow support arm and concrete matrix are all clearly visible in the surrounding air.  These images are described in detail in the text.}}
\label{fig:Uimage}
\end{figure*}

\begin{figure*}[t] 
\centering 
\includegraphics[width=1.0\columnwidth,keepaspectratio]{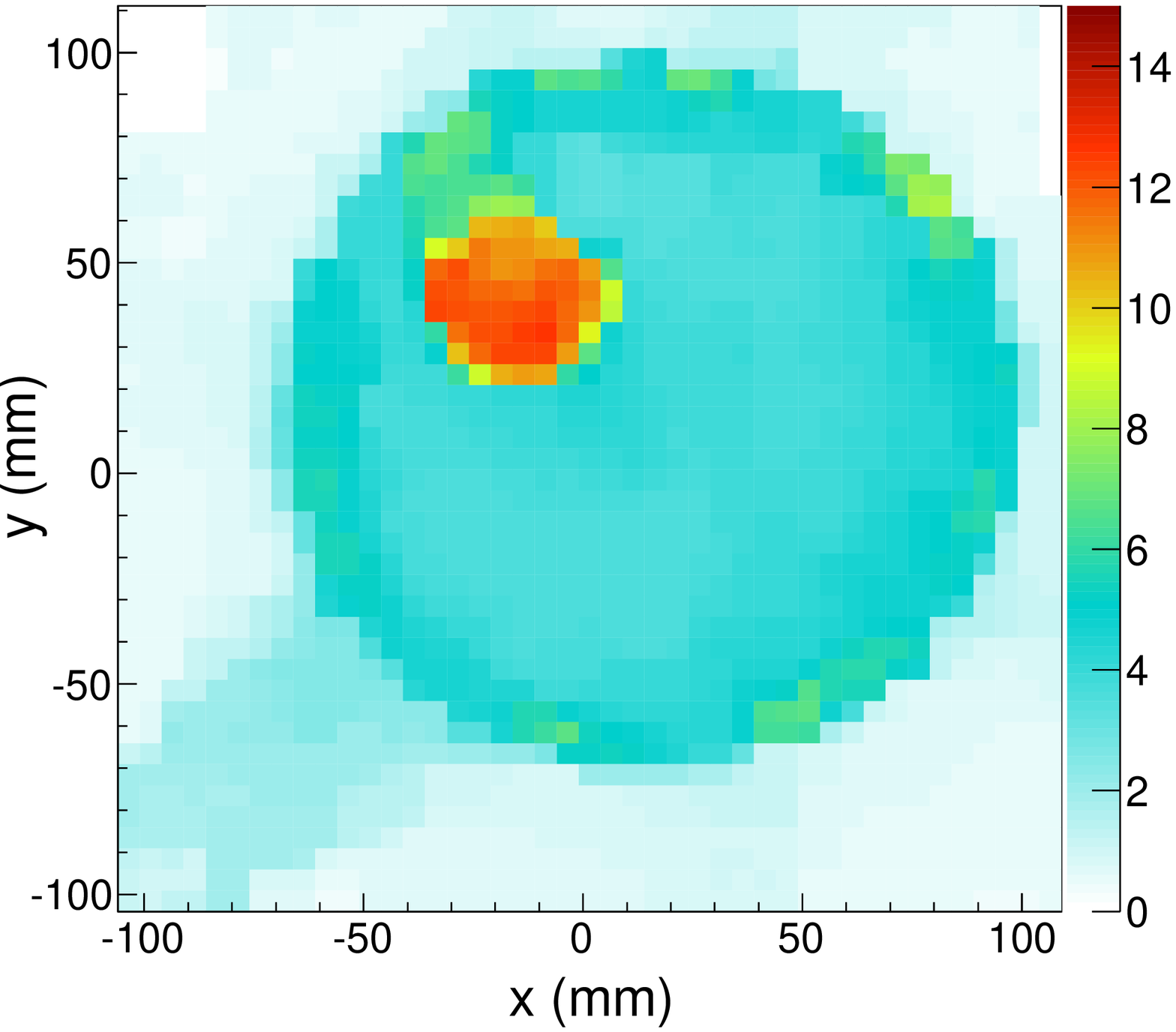}
\includegraphics[width=1.0\columnwidth,keepaspectratio]{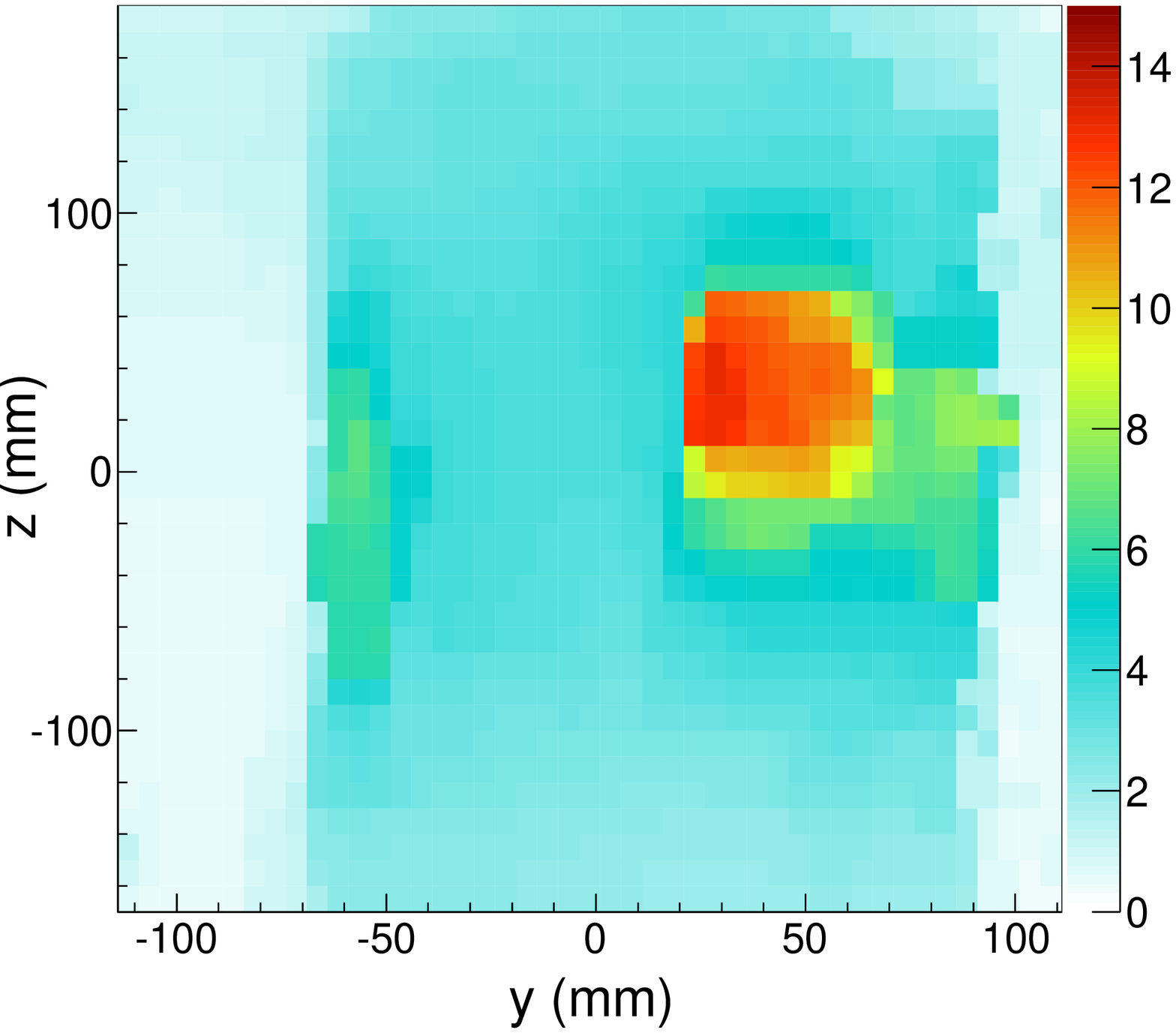}
\caption{\small{\textbf{Tomograms through the lead-containing region of the barrel.}  Images reconstructed from several weeks of exposure to cosmic-ray muons using the small-scale prototype system.  Shown are a 10\,mm slice in the xy-plane \ie horizontal through the assay volume (left) and a 5\,mm slice in the vertical $yz$-plane (right) through the region containing the lead cube.  Here, the colour scale denotes the most-likely $\lambda$ value in each voxel reconstructed by the image reconstruction software.  Only voxels within the active volume are shown.  The cube of lead, stainless steel casing, concrete matrix and smearing effects from the support arm are all clearly visible in the surrounding air.  These images are described in detail in the text.}}
\label{fig:Pbimage}
\end{figure*}

Images reconstructed from experimental data collected using this industrial test configuration are presented in Figures~\ref{fig:Uimage}~and~\ref{fig:Pbimage} for slices (or tomograms) through two orthogonal planes containing the uranium and lead samples respectively.    In both of the 10\,mm tomograms through the xy plane \ie the horizontal slices, the 12\,mm-thick circular contour of the stainless-steel barrel and support arm are clearly visible against the background of the surrounding air.   This tomogram in Figure~\ref{fig:Uimage} shows the cylindrical sample of uranium reconstructed in the centre of the concrete matrix with a $\lambda$ value in the region of 14\,mrad$^{2}$\,cm$^{-1}$.  The halo of voxels with lower $\lambda$ values of 9\,mrad$^{2}$\,cm$^{-1}$ around the uranium object represent voxels containing partial contributions from both uranium and concrete, which acted to dilute the reconstructed scattering density.   In this horizontal tomogram the air void at the centre of the support arm is also detected successfully.

The 5\,mm vertical slice through the assay volume shown in Figure~\ref{fig:Uimage} again highlights the clarity of the uranium image obtained within the concreted container.   Characteristic smearing of the image in the z axis was observed.  This is an inherent effect introduced in the principle axis of muon momentum (\ie the z direction) from the reconstruction of the scattering location for milliradian-order scattering.  Here, an approximate increase of 30\% in the length of this cylinder was observed.  The 12\,mm-thick stainless-steel casing is visible at the edges of the uniform region (with $\lambda$ values of around 4\,mrad$^{2}$\,cm$^{-1}$) of concrete.  The success in reconstructing a uranium cylinder of these dimensions, comparable with nuclear fuel elements, further confirms the potential of this technology for the characterisation of legacy nuclear waste.

Similar results were observed for the corresponding tomograms through the centre of the lead-containing region.    In Figure~\ref{fig:Pbimage} the lead cube was observed in the xy plane at an angle of 45\degree{} close to the barrel wall with $\lambda$ values in a small range from 11 to 15\,mrad$^{2}$\,cm$^{-1}$.  The dimensions of the high-Z lead cube in this plane were reconstructed to millimetre precision.  Again, across the concrete material, a uniform $\lambda$ distribution of approximately 4\,mrad$^{2}$\,cm$^{-1}$ was reconstructed within the steel contours.  In the accompanying slice through the yz plane the vertical smearing was observed to roughly the same extent as shown in Figure~\ref{fig:Uimage}.   The reconstructed $\lambda$ values for both of the high-Z materials appear suppressed in comparison with those determined within an air matrix from Ref.~\cite{Clarkson2014a}.   This effect is discussed in the following. 

\begin{figure}[t] 
\centering 
\includegraphics[width=1.0\columnwidth,keepaspectratio]{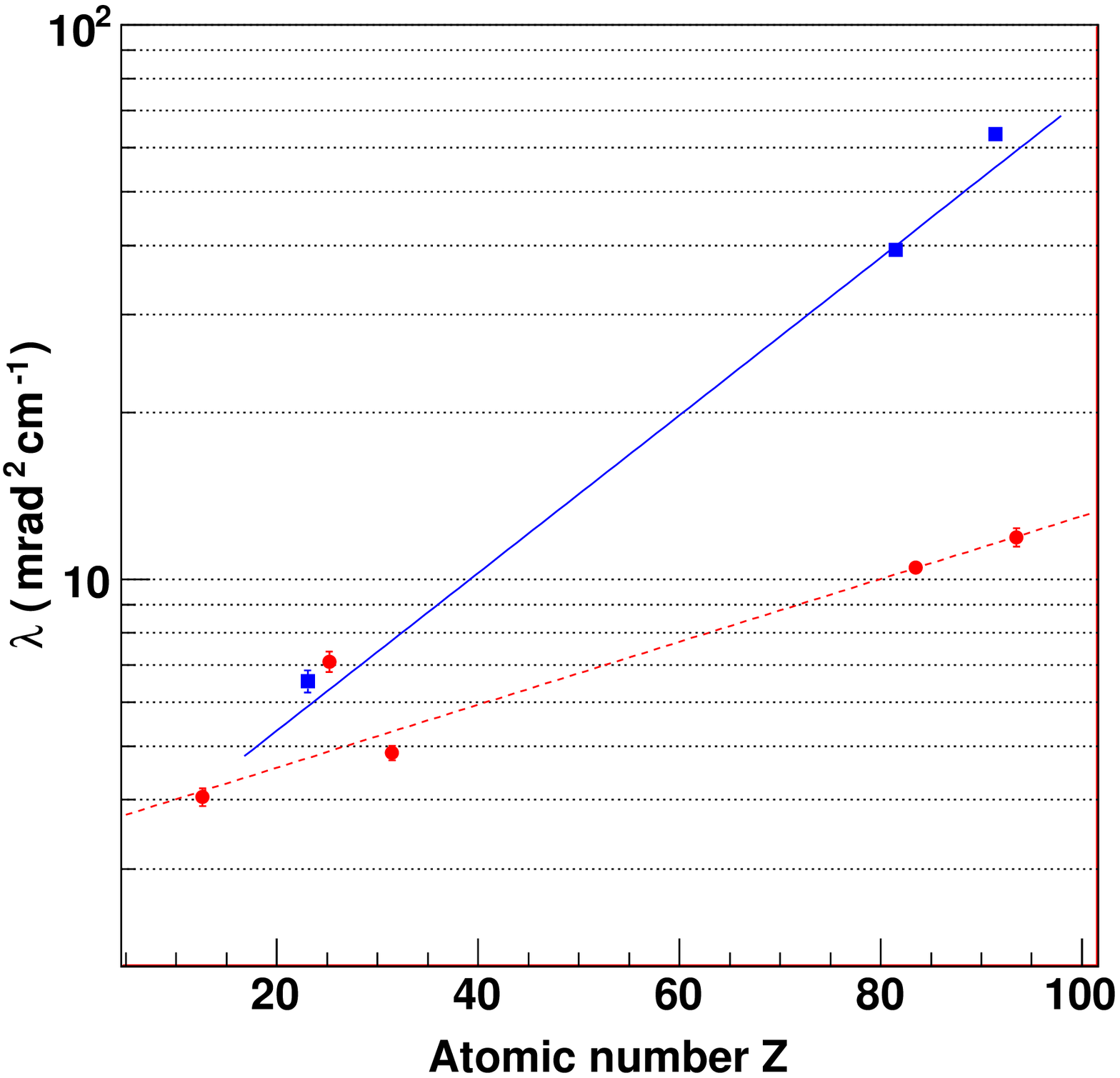}
\caption{\small{\textbf{Material discrimination comparison in air and concrete matrices.}  Semi-logarithmic comparison between average $\lambda$ values reconstructed from experimental data collected for a range of different materials within air (blue squares) and concrete (red circles) matrices.  From left to right, the materials tested are concrete, stainless steel, brass, lead and uranium.  Data have been extracted from Figures~\ref{fig:Uimage}~and~\ref{fig:Pbimage} and from updated results on an experimental setup first presented in Ref.~\cite{Clarkson2014a} using the same high-Z objects.  Shown are exponential fits to the data.  This result is described in detail in the text.}}
\label{fig:AirConComp}
\end{figure}

In Ref.~\cite{Clarkson2014a}, our first results were reconstructed using the same uranium and lead samples with data collected over a similar timescale within an air matrix.   The average $\lambda$ values in the regions identified as containing these high-Z materials within air and concrete matrices are compared in Figure~\ref{fig:AirConComp}.   The suppression of the reconstructed $\lambda$ values for uranium and lead when observed within concrete is clearly observed.  This is attributed to the increased Coulomb scattering contribution from the concrete in the path of the muon that effectively reduces the observed scattering contribution from the high-Z materials. However, from the tomograms presented in Figures~\ref{fig:Uimage}~and~\ref{fig:Pbimage}, the required discrimination between the high-Z materials and the concrete matrix mandated by this industrial scenario was comfortably achieved.   In both the air and concrete data sets, the extracted $\lambda$ values have also been shown to exhibit exponential dependences on Z, thus providing clearer discrimination for such materials.

Figure~\ref{fig:AirConComp} also presents the result extracted from tomograms through the region of the concreted brass, shown with an effective Z of approximately 30.   This sample was observed with an average $\lambda$ of approximately 20\% above the level of the surrounding concrete in a region consistent with the known location within the container.  The result extracted from the external stainless-steel casing (with effective Z of approximately 24) of the concrete-filled barrel is, as expected, in agreement with the value extracted from the same material observed in air.

\section{Summary}\label{summary}
First imaging results from experimental data collected using a prototype detector system have been presented for an industrial test configuration of low- and high-Z materials encapsulated within a small concrete-filled barrel. The images accurately reflect the interior of the 12\,mm-thick stainless steel barrel and demonstrate the ability to investigate sealed containers with high spatial precision.  The clear discrimination between the high-Z materials and the concrete matrix, and the precision spatial resolution confirm that MT is certainly applicable to aspects of nuclear waste reprocessing, where the ability to investigate sealed containers is highly desirable.  It is therefore foreseen that with further research, and the development of a full-scale system that will considerably improve resolution and reduce the required exposure times, this technology will be employed within the UK Nuclear Industry to help mitigate the risks inherent with long-term storage of nuclear waste.

\section*{Acknowledgements}
The authors gratefully acknowledge Sellafield Ltd., on behalf of the UK Nuclear Decommissioning Authority, for their continued support and funding of this project.

\end{document}